\documentclass[preprint]{aastex631}

\newcommand{\history}[1]{} %
\newcommand{\Msun}[1]{M$_\odot$}
\begin{document}

\title{The Periastron Passage of T Tauri South B as Viewed by ALMA: Millimeter Flux Variations and Dust Heating Triggered by Orbital Motion} 

\author[0000-0002-6881-0574]{Tracy L. Beck}
\affiliation{Space Telescope Science Insitute, 3700 San Martin Drive, Baltimore, MD USA}
\author[0000-0003-3773-1870]{Stephane Guilloteau}
\affiliation{Laboratoire d’Astrophysique de Bordeaux, Université de Bordeaux, CNRS, B18N, Allée Geoffroy Saint-Hilaire, F-33615 Pessac, France}
\author[0000-0001-5415-9189]{Gail Schaefer}
\affiliation{The CHARA Array of Georgia State University, Mount Wilson Observatory, Mount Wilson, CA 91023, USA}
\author[0009-0003-6480-8084]{Edwige Chapillon}
\affiliation{Laboratoire d’Astrophysique de Bordeaux, Université de Bordeaux, CNRS, B18N, Allée Geoffroy Saint-Hilaire, F-33615 Pessac, France} 
\affiliation{The Institut de Radioastronomie Millimétrique (IRAM), 300 rue de la Piscine, F-38406 Saint-Martin d’Héres, Grenoble, France}
\author[0000-0003-2341-5922]{Anne Dutrey}
\affiliation{Laboratoire d’Astrophysique de Bordeaux, Université de Bordeaux, CNRS, B18N, Allée Geoffroy Saint-Hilaire, F-33615 Pessac, France}
\author[0009-0009-9618-4927]{Emmanuel Di Folco}
\affiliation{Laboratoire d’Astrophysique de Bordeaux, Université de Bordeaux, CNRS, B18N, Allée Geoffroy Saint-Hilaire, F-33615 Pessac, France}



\begin{abstract}
We present 225 and 350~GHz imaging of the iconic T Tauri system using the Atacama Large Millimeter submillimeter Array (ALMA).  T Tauri is a hierarchical triple system, and the close binary T~Tau~Sa/Sb underwent periastron passage in March 2023.  The ALMA images were obtained in epochs spanning November 2019 through June 2023, and therefore covered the time frame of the recent periastron passage.  We clearly resolve the Sa-Sb binary in two epochs of high-resolution measurements with ALMA.  We find increases in millimeter flux from heating of the Sa disk and the wider distribution of dust in the environment of the binary.  This heating is likely in response to increased stellar accretion activity triggered by orbital motion during the dynamic periastron passage of T Tau Sb around Sa.  Resolved, extended millimeter emission is also found to change morphology and increase in flux in the immediate environment of the Sa-Sb binary after periastron passage.  This may suggest an increase in nonthermal emission from magnetic interaction, gravitational disruption of the circumstellar disks as the stars passed through periastron, or both of these phenomena. We also detected structures in the compact (24~au radius), thermal dust disk around T~Tau~N. In particular,  we identify a crescent-shaped emission excess just outside a shallow gap at 12 au radius that appears to move at Keplerian speed.  Future measurement of dust spectral indices can clarify the origin of increased and variable millimeter emission in the environment of the T~Tau~S binary.

\end{abstract}



\section{Introduction} \label{sec:intro}

The T Tauri system defined its class of low mass sun-like stars \citep{joy45}. Yet, detailed multi-wavelength observations at increasing spatial resolution have revealed that T Tauri is a remarkable young triple star. T Tau South (T Tau S), the southern infrared companion, was discovered in 1980 \citep{dyck82}. In 1997, T Tau S was spatially resolved into two stars, Sa and Sb \citep{kore00}. This $\sim$0$\farcs$1 separation T Tau S binary is obscured to optical invisibility by circumbinary material, even though it is located only at a projected 0$\farcs$7 distance to the south of the bright optical star, T Tauri North \citep[T Tau N;][]{beck20}. The motion of the Sa+Sb binary system has now been monitored through most of its orbit \citep{kohl16, scha20}. The best-fit orbital period is 27 years and the most recent binary periastron closest approach occurred in March 2023 \citep[2023.3±0.4;][]{scha20}.

The extended mass outflow known as Burnham's variable nebula \citep{burn90} extends $\sim$10$''$ to the south of T Tau; it has varied in morphology throughout the past century and it has an optical spectrum typical of Herbig-Haro outflows \citep{herb50}. This southern flow was spatially resolved into loops and arcs of H$\alpha$ emission \citep{robb95}. Interestingly, these arcs and loops of emission seen in the extended southern Burnham's nebula outflow seem to be periodic, which has led to the intriguing possibility that the extended outflows were triggered by orbital motion of the T Tau Sa+Sb binary \citep{beck20}. 
Although rare and poorly understood, mass outflows triggered by binary orbital motion have been postulated to explain regular outflow ejection in some systems \citep[e.g., HH30;][]{angl07}. The binary periastron passage is theorized to disturb inner disks in a rapid and enhanced mass accretion, causing subsequent mass outflow. Enhanced mass accretion from binary orbital motion is well established for pulsed-accreting binaries such as DQ Tau and LRLL 54361  
\citep{basr97, muze13}.  The T Tau S system experiences such accretion pulses on a longer time frame because of the wider orbit \citep{beck20}.  As such, T Tau S is a prime laboratory for investigating changes that may be triggered by close periastron passage from the binary orbital motion. 

The millimeter wavelength continuum emission from T Tau has 4 main components \citep{beck20}: 1) a spatially resolved circumstellar disk around T Tau N, 2) a spatially resolved disk around T Tau Sa of size $\sim$7 au diameter, elongated north-south, 3) a spatially unresolved disk around T Tau Sb, and 4) an extended, elongated emission component that is interpreted as dust in the circumbinary disk around the T Tau S system \citep[Figure 4 in][]{beck20}. The resolved circumstellar disks around T Tau N and Sa are noncoplanar to each other, and also misaligned with the circumbinary disk geometry \citep{beck20}. Remarkably, the $\sim$7\,au diameter resolved disk around T Tau Sa extends a substantial fraction of the Sa-Sb binary periastron distance ($\sim$5.5\,au). Therefore, some extreme tidal disk interaction effects may be expected during binary periastron passage.

T Tau Sa and Sb now provide the unique and exciting opportunity to measure multi-wavelength disk, stellar accretion and outflow properties, while also resolving the binary through periastron passage in anticipation of a new mass outflow ejection.  With its ejection phenomena correlated with its periastron passage, evidence for three misaligned circumstellar disks and at least two misaligned outflows \citep{beck20}, T Tau offers an important opportunity to witness and map in real-time the tidal striping effects, dynamically enhanced accretion, and accretion outburst-triggered mass outflow ejection in this emblematic young triple system.
In this study, we present two epochs of moderate (0$\farcs$7) and high (0$\farcs$03-0$\farcs$05) resolution mapping of the mm continuum emission from the T Tauri environment, as measured by ALMA.  The observations span the 2019 - 2023 time frame, and hence map the mm emission as T Tau Sb was moving extremely rapidly around Sa through the orbital periastron passage.  We analyze the maps of millimeter emission and changes in morphology, and where possible variations in the emission spectral index.  We seek to understand the dynamic and complex interactions of material, the energetic processes associated with rapid orbital motion in the Sa - Sb system, and discuss interpretations in this context.

\section{Observations}
The main observations for this project were acquired in ALMA Cycle 7 and Cycle 9, for a two-epoch dataset (program IDs 2019.1.00703.S, and 2022.1.01134.S, respectively).  High-resolution observations in Cycle 7 were delayed by one year because of the pandemic shutdown. As a result, the two epochs of observations at each of three array configurations were separated in time by about two years. Table \ref{tab:obs} provides the details of the sets of observations presented in this study. 

In both epochs, three sets of data were acquired with ALMA to measure the mm dust continuum components in the T Tau system:\\

A) High resolution Band 6 – 0$\farcs$03 to 0$\farcs$05: continuum in the circumstellar disk components and Sa–Sb orbit: The very high angular resolution observation with ALMA at long baselines, 30 mas resolution, was required 1) to establish the disks flux prior to the periastron passage, and 2) to trace the binary motion in complement to IR measurements. 

B) Band 7 – 0$\farcs$15 - disks and extended continuum:  a 0$\farcs$15 resolution was sufficient to separate Sa and Sb by model fitting (with known positions from Band 6 data). This data also included sufficient number of short baselines to provide sensitivity to the extended dust, such as associated with the circumbinary disk around T Tau S.

C) Medium resolution Band 6 – 0$\farcs$7 - extended continuum:  Search at higher sensitivity for emission from dust in the circumbinary disk and at lower resolution (wider scale). 

The spectral index is measured between Bands 6 and 7, or using the frequency difference between the Upper and Lower sidebands in Band 6,
to discriminate between thermal and non-thermal contributions.

\begin{table}
\begin{tabular}{c|c|c|l|c}
\hline\\
UT Date & Band & Approximage Frequency  & Approximate   & Note \\
 		&  &  Coverage & Angular Resolution   & \\
\hline
\hline
 \multicolumn{5}{c}{Epoch 1 - ALMA Cycle 7  2019.1.00703.S}\\
\hline
26-Nov-2019 & 7 & [342-346] [354-357] & C43-3 (0.70$''$) & Low resolution 2019 \\ 
21-Dec-2019 & 7 & [342-346] [354-357] & C43-3 (0.70$''$) & Low resolution 2019 \\ 
13-Jun-2021 & 6 & [216-221] [230-235] & C43-6 (0.25$''$) & High resolution 2021 \\ 
28-Aug-2021 & 6 & [216-221] [230-235] & C43-9 (0.03$''$) & High resolution 2021\\
09-Nov-2021 & 6 & [216-221] [230-235] & C43-7 (0.25$''$) & High resolution 2021\\
 \multicolumn{4}{c}{Epoch 2 - ALMA Cycle 9 2022.1.01134.S}\\
\hline
01-Oct-2022	& 7 & [342-346] [355-356] & C43-2 (1.00$''$) & Low resolution 2022 \\
20-Oct-2022 & 6 & [217-221] [230-235] & C43-2 (0.70$''$) & Low resolution 2022 \\
25-Jun-2023 & 6 & [216-221] [230-235] & C43-8 (0.05$''$) & High resolution 2023 \\

\hline
\end{tabular}
\caption{Observing dates and approximate angular resolution. 
Given the source variability, we used only data sufficiently
close in time in this study.}
\label{tab:obs}
\end{table}

\section{Data Analysis} \label{sec:style}
\label{sec:analysis}

All of the ALMA data was initially calibrated with CASA, using the scripts and calibration parameters from the ALMA archive, and then exported through \texttt{exportuvfits} into UVFITS format.
Further processing was done using the \textsc{IMAGER}\footnote{https://imager.oasu.u-bordeaux.fr} program.
The data were shifted using proper motions
from GAIA and referenced to T Tau North position at Epoch 2000.0 \citep[04:21:59.4319 +19:32:06.4324][]{gaia22}.

Imaging was performed using the \textsc{IMAGER} built-in pipeline. The pipeline derives self-calibration parameters from wide band spectral windows, and applies them as pathlength corrections to all simultaneously available
spectral windows.  Continuum and spectral lines are separated
using a simple heuristic using a sigma clipping method that assumes constant continuum flux over each spectral window.   After masking the line contaminated regions, multi-frequency synthesis has been performed on the line-free continuum data, using a spectral index of 2 for the emission, as justified by the analysis presented hereafter in Section \ref{sec:sub:flux}. 
  
\begin{figure}
\includegraphics[width=18.0cm]{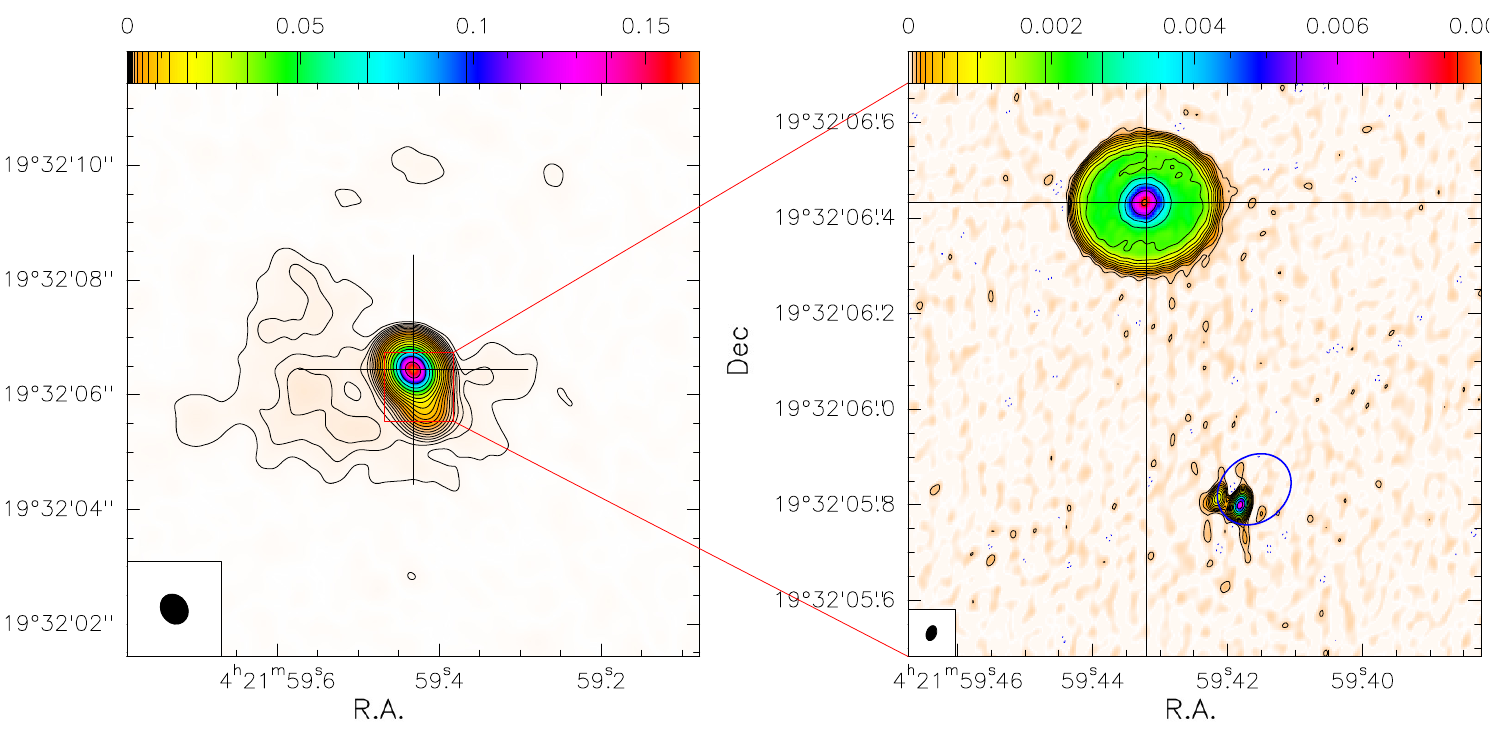}
\caption{
(Left) Low resolution image of T Tau at 225 GHz in 2021, showing
the low level extended emission. Noise level is 45 $\mu$Jy/beam, 4.2 mK at this resolution of $0.55 \times 0.47''$  at PA  $32^\circ$.
Contour levels are -12, 12, 24, and 36\,mK and then in steps
of $\sqrt{2}$ up to 13\,K. 12\,mK is about 128 $\mu$Jy/beam or $2.8 \sigma$.
\history{Figure produced by ANALYSIS/230GHz/whole-2021.ima}
(Right): High resolution image of T Tau at 225 GHz in 2021.
Noise level is 20 $\mu$Jy/beam, 0.7\,K at 
at this resolution of $32.7 \times 20.9$ mas at PA  $160^\circ$.
Contour levels are -2, 2, 4, and 6\,K and then in steps
of $\sqrt{2}$ up to 256\,K. 2\,K is about 57 $\mu$Jy/beam or $2.8 \sigma$.
The cross marks the GAIA position at the epoch of observation
(Nov 2021). The blue ellipse presents the best fit orbital model
for Sa, Sb, using orbital parameters from \citet{scha20}.  Beam sizes are presented at lower left for clarity. The color scale indicates the flux density in mJy/beam.
\label{fig:fig1}}
\end{figure}

Given the large spread of baseline lengths, for source decompositions we focus primarily on the longest baselines 225~GHz data, which have been obtained on two epochs; 2021 and 2023.  Observations for both epochs use different array configurations and provide somewhat different angular resolution. The low resolution (left) and high resolution (right) images, obtained in Nov 2021 and Aug 2021, respectively, are shown in Fig.\ref{fig:fig1}, while Fig.\ref{fig:fig2} shows similar images for Jun 2023.
For further comparison between epochs, we shall use images built at the same angular resolution, through a combined use of weighting, tapering, and convolution of the Cleaned images to obtain a circular beam of 42.5 mas FWHP \citep[6.2 au at the distance of T Tau, $145.1 \pm 0.8$ pc from GAIA,][]{bail18}. The 350 GHz data were treated in a similar way, but only provide a much more moderate angular resolution. Since the noise heavily depends on the weighting function, the noise levels are indicated in the figure captions for all images.

\subsection{Decomposition in sources}
\label{sec:sub:sources}

\begin{figure}
\includegraphics[width=18.0cm]{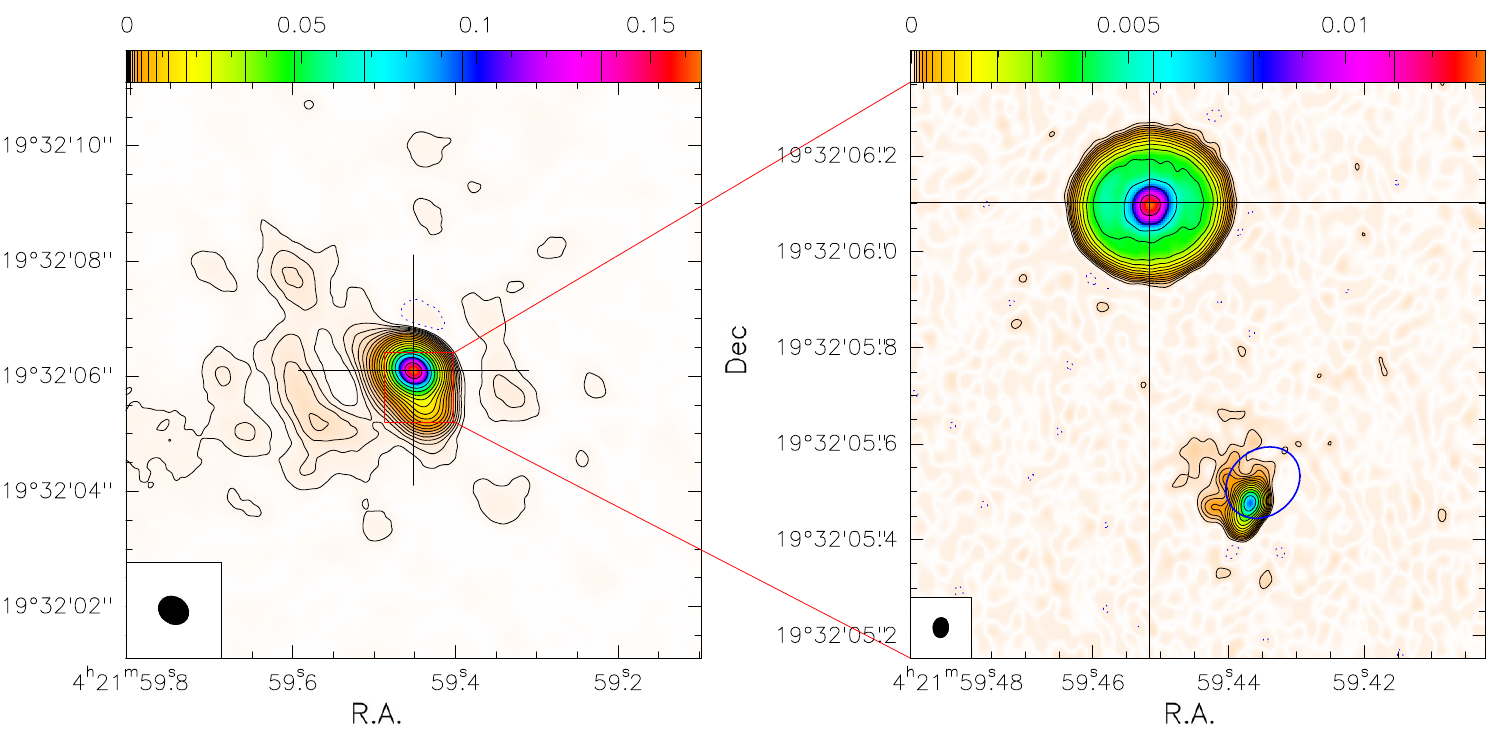}
\caption{As in Figure 1, but for the 2022-2023 data. Left: low resolution image combining Oct 2022 and June 2023 data.  Noise level is 59 mJy/beam, 5.5 mK at the angular resolution of $0.55 \times 0.47''$ at PA $56^\circ$. Contour spacings
are 16.5, 33, 48.5 mK and then in steps of $\sqrt{2}$ up to 12.7\,K.
Right: High resolution image using June 2023 data only. Noise level is 19 mJy/beam, 0.33\,K at the angular resolution of $0.043\times 0.033''$ at PA $174^\circ$. 
Contour levels are -1, 1, 2, and 3\,K and then in steps of 
$\sqrt{2}$ up to 128\,K.
A 25 mas shift in Declination was required to register both data sets.  The blue ellipse presents the best fit orbital model
for Sa, Sb, using orbital parameters from \citet{scha20}.  In the high resolution image (right), the position of T Tau Sb has changed because of orbital motion from East-Northeast in 2021 (Figure 1) to South-Southeast here in 2023.  Beam sizes are presented at lower left for clarity.
\label{fig:fig2}}
\end{figure}

The continuum emission around T Tau is made of 5 components:
the disk around T Tau N, the disk around T Tau Sa, a compact
source coincident with T Tau Sb, and an extended (of order 2$''$)
emission, mostly East-South-East of T Tau N, that is only
visible in the 350 GHz data given its average surface brightness (see \S 4.2). In addition, a new, spatially resolved but compact (scale of order 0.1$''$), emission region appears around T Tau Sa/b in the June 2023 data (see Fig.\ref{fig:fig2}), i.e. near the epoch of the peri-astron passage of the Sa/Sb system. 

The disk of T Tau N itself was decomposed at first order into a nearly uniform disk, to which a compact central Gaussian shaped source is superimposed. Using the highest resolution data from Nov 2021, Table \ref{tab:ttaun} summarizes the fit results of such a decomposition. The fit was made in the $uv$ plane, after removal of the best fit Gaussians for Sa and Sb. This disk fit quality was investigated by looking at the azimuthally averaged visibility profile after deprojection from the PA and inclination $i$ derived from the uniform disk component, and by imaging the residuals (see Section \ref{sec:ttaun} for further discussion).
%

The source properties determined from the decompositions for N, Sa and Sb, in particular integrated fluxes, are given in Table \ref{tab:sources}. Sa and Sb results were obtained by simultaneous Gaussian fit to the $uv$ data after removal of the Clean components located around T Tau N.  In both cases in Tables \ref{tab:ttaun}, \ref{tab:sources} and \ref{tab:sources2023}, the $\Delta\alpha$ and $\Delta\delta$ values represent the offset between the millimeter flux position and the GAIA reference position of T Tau N at the epoch of the observation.

\subsection{Calibration consistency \& Spectral indexes}
\label{sec:sub:flux}
T Tau N, whose structure does not exhibit any prominent change between
Aug-2021 and Jun-2023, can be used to verify the calibration consistency
between the two epochs. The flux ratio (2023 / 2021) is $0.987 \pm 0.001$
for both the peak flux and the total source flux at both 217 GHz and 234 GHz.
The flux of T~Tau~N is consistent at the 1.5\% level across the two epochs, this is 
well below the estimated absolute flux calibration accuracy of ALMA. 

In Aug 2021, the achieved sensitivity and angular resolution allow
a sufficiently precise determination of the spectral index of 
the emission from T Tau N, Sa and Sb by comparing flux at 217 GHz and 234 GHz. 
N and Sa both have a spectral index of 2.0  (Table \ref{tab:sources}), consistent with optically
thick thermal emission from a dusty disk.  However, emission from Sb exhibits a spectral
index below 0 ($-0.7 \pm 0.3$), which is not consistent with thermal dust emission and likely arises from non-thermal processes (see also \S 5).  There is a relatively low frequency difference in the measurement
and fainter signal, resulting in a high uncertainty.  However, the spectral index of the emission from Sb does not appear to be consistent at a level greater than 5$\sigma$ with optically thick thermal emission predominantly arising from a circumstellar disk.  Optically thin dust emission is ruled out at an even higher significance level.
\begin{table}
\begin{tabular}{l|c|c}
\hline
 & \multicolumn{2}{c}{T Tau N} \\
 & {Uniform Disk} &  {Gaussian} \\
\hline
$\Delta\alpha$ (mas) & $-4.64\pm0.02$  & $4.73\pm0.03$ \\    
$\Delta\delta$ (mas) & $-2.38\pm0.1$  & $-1.36 \pm 0.03$ \\
Flux (mJy)      & $147.94 \pm 0.06$ & $31.52 \pm 0.05$ \\
Major (mas)     & $288.7 \pm 0.1$ & $63.5 \pm 0.1$ \\
Minor (mas)     & $252.1 \pm 0.1$ & $50.6 \pm 0.1$ \\
PA  ($^\circ$) & $93.6 \pm 0.1$ & $ 96.3 \pm 2$ \\
i   ($^\circ$) & $29.2 \pm 0.1$ & $37.3 \pm 0.2$ \\
\hline
\end{tabular}
\caption{Decomposition in two components of T Tau North continuum
emission at 225 GHz. $i$ is the inclination assuming the aspect
ratio is due to projection effect of a circular structure.
\label{tab:ttaun}}
\end{table}

\begin{table}
\begin{tabular}{l|c|c}
\hline
 & \multicolumn{2}{c}{T Tau N} \\
 \hline
 & {216.8 GHz} &  {234.1 GHz} \\
Flux (mJy) & $169.5\pm0.2$ &  $198.8 \pm 0.4$ \\
Spectral index   & \multicolumn{2}{c}{$2.08 \pm 0.04$} \\
\hline
 & \multicolumn{2}{c}{T Tau Sa} \\
 \hline
 & {216.8 GHz} &  {234.1 GHz} \\
$\Delta\alpha$ (mas) & $-195.2\pm0.1$ & $-195.4 \pm 0.1$ \\    
$\Delta\delta$ (mas) & $-633.5\pm0.1$ & $-633.3 \pm 0.1$ \\
Flux (mJy)      & $6.75 \pm 0.02$ & $7.87 \pm 0.02$ \\
Major (mas)     & $14.4 \pm 0.4$ & $14.5 \pm 0.4$ \\
Minor (mas)     & $ 8.4 \pm 0.4$ & $8.9 \pm 0.3$ \\
PA  ($^\circ$) & $159 \pm 2$ & $ 161 \pm 2$ \\
Spectral index   & \multicolumn{2}{c}{$2.00 \pm 0.07$} \\
\hline
 & \multicolumn{2}{c}{T Tau Sb} \\
\hline
 & {216.8 GHz} &  {234.1 GHz} \\
$\Delta\alpha$ (mas) & $-148.4\pm0.3$  &  $-148.3\pm0.3$  \\
$\Delta\delta$ (mas) & $-617.9\pm0.3$  & $-618.7 \pm 0.3$ \\
Flux (mJy)      & $1.48 \pm 0.02$ & $1.40 \pm 0.02$ \\
Major (mas)     & $19 \pm 2$ & $15\pm 2$ \\
Minor (mas)     & $10 \pm 2$ & $6\pm 2$ \\
PA  ($^\circ$) & $173 \pm 6$ & $140 \pm 8$ \\
Spectral index   & \multicolumn{2}{c}{$-0.72 \pm 0.36$} \\
\hline
\end{tabular}
\caption{Properties of the continuum emission around N, Sa and Sb
in August 2021.}
\label{tab:sources}
\end{table}

\begin{table}
\begin{tabular}{l|c|c}
\hline
\hline
 & {T Tau Sa} & T Tau Sb \\
\hline
$\Delta\alpha$ (mas) & $-205.5\pm0.1$ & $-187.9 \pm 0.2$ \\    
$\Delta\delta$ (mas) & $-617.6\pm0.1$ & $-654.9 \pm 0.3$ \\
Flux (mJy)      & $8.96 \pm 0.02$ & $2.08 \pm 0.04$  \\
Major (mas)     & $20.1 \pm 0.5$ & $13.1 \pm 0.9$ \\
Minor (mas)     & $16.0 \pm 0.2$ &  [2] \\
PA  ($^\circ$)  & $148 \pm 3$ &  $66 \pm 8$ \\
\hline
\end{tabular}
\caption{Properties of the 217 - 234 GHz continuum emission around Sa and Sb
in Jun 2023. The minor axis of Sb is unresolved, and was fixed to 2 mas
(the estimated upper limit at the $3\sigma$ level is about 10 mas).
} 
\label{tab:sources2023}
\end{table}

\section{Results} \label{sec:results}

\subsection{225 GHz Measurements}
\label{sec:sub:images}
\begin{figure}
\begin{center}
\includegraphics[width=17.0cm]{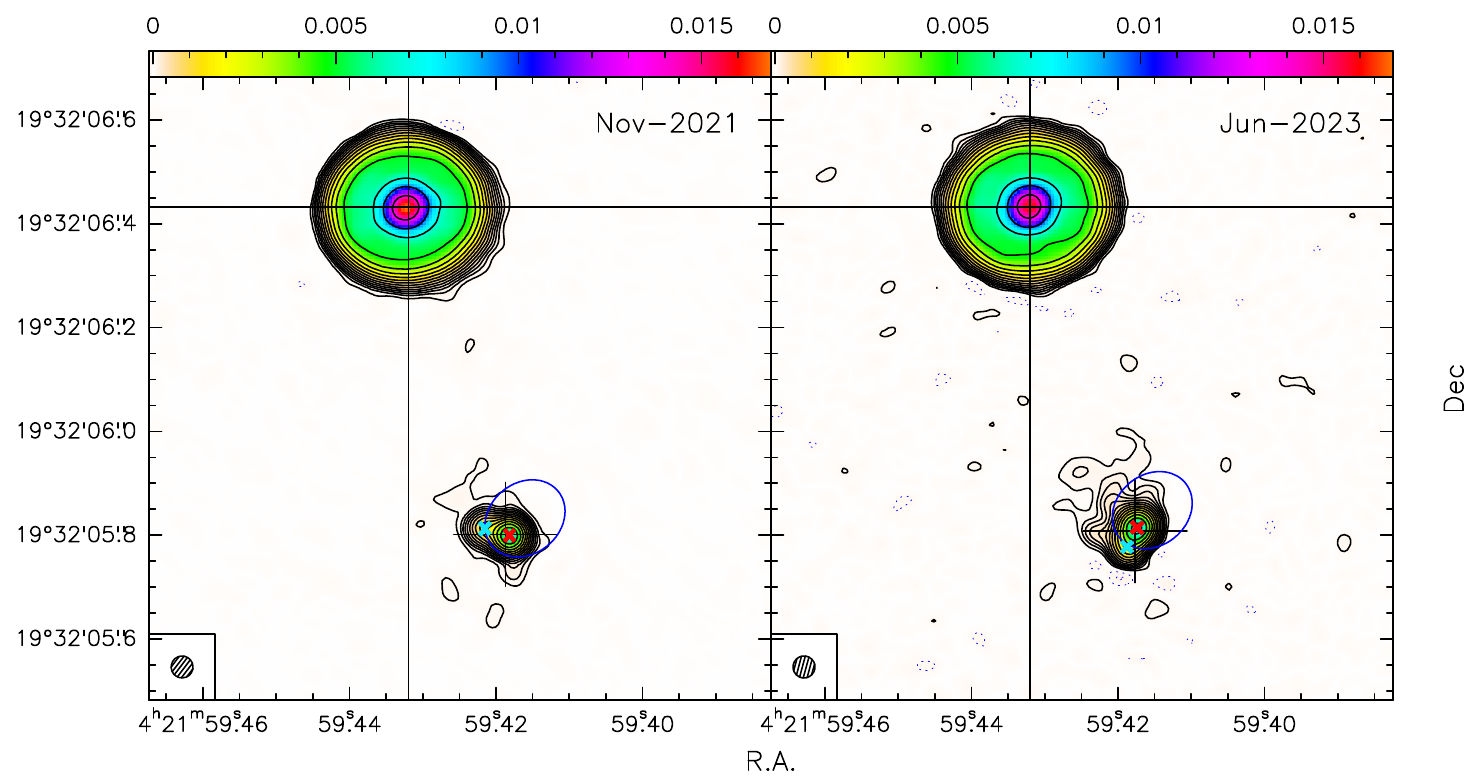}
\caption{A comparison of the 225 GHz continuum emission at the
two epochs (09-Nov-2021, 23-Jun-2023), at the same 42.5 mas spatial
resolution. Contour
levels are in steps of 0.666 K (about 50 $\mu$Jy/beam, approximately 4.7\,$\sigma$
in Nov-2021 and 3.1\,$\sigma$ in Jun-2023 respectively), up to 2 K and then 
increase by a factor $\sqrt{2}$ up to 128 K (peak value 225 K).
The red and cyan X's mark the measured Sa and Sb positions in the two epochs, and the black cross is the Sa/Sb center of mass. Beam sizes are presented at lower left for clarity.
\label{fig:fig3} }
\end{center}
\end{figure}

We combined the 217 and 234 GHz continuum data described in the previous section, \ref{sec:sub:flux}, using an average spectral
index of 2 and a reference frequency of 225 GHz, which optimizes the sensitivity for most emission, without biasing the flux of Sb despite its different spectral index. Fig.\ref{fig:fig3} shows the comparison of this 225 GHz continuum emission at the two epochs, at the same angular resolution. No position adjustment were made, the phase center remains on the position for T Tau N predicted from the GAIA measurements (marked by the perpendicular black lines across the field). The red and cyan X's mark the position of T Tau Sa and Sb, obtained from the estimated orbit in 2021, and an extrapolated position for 2023.

This position estimate for 2023 is made assuming that the peak of 1.3\,mm emission is centered on Sa, and there is a separation of 35 mas at an angle of $160^\circ$ between Sa and Sb from prior orbital projections \citep{scha20}. The black plus is the position of the center of mass of the Sa/Sb system, using stellar masses of 
2.03, 2.05 and 0.43 \Msun{} for N, Sa and Sb
\citep{scha20}.

\begin{figure}
\includegraphics[width=18.0cm]{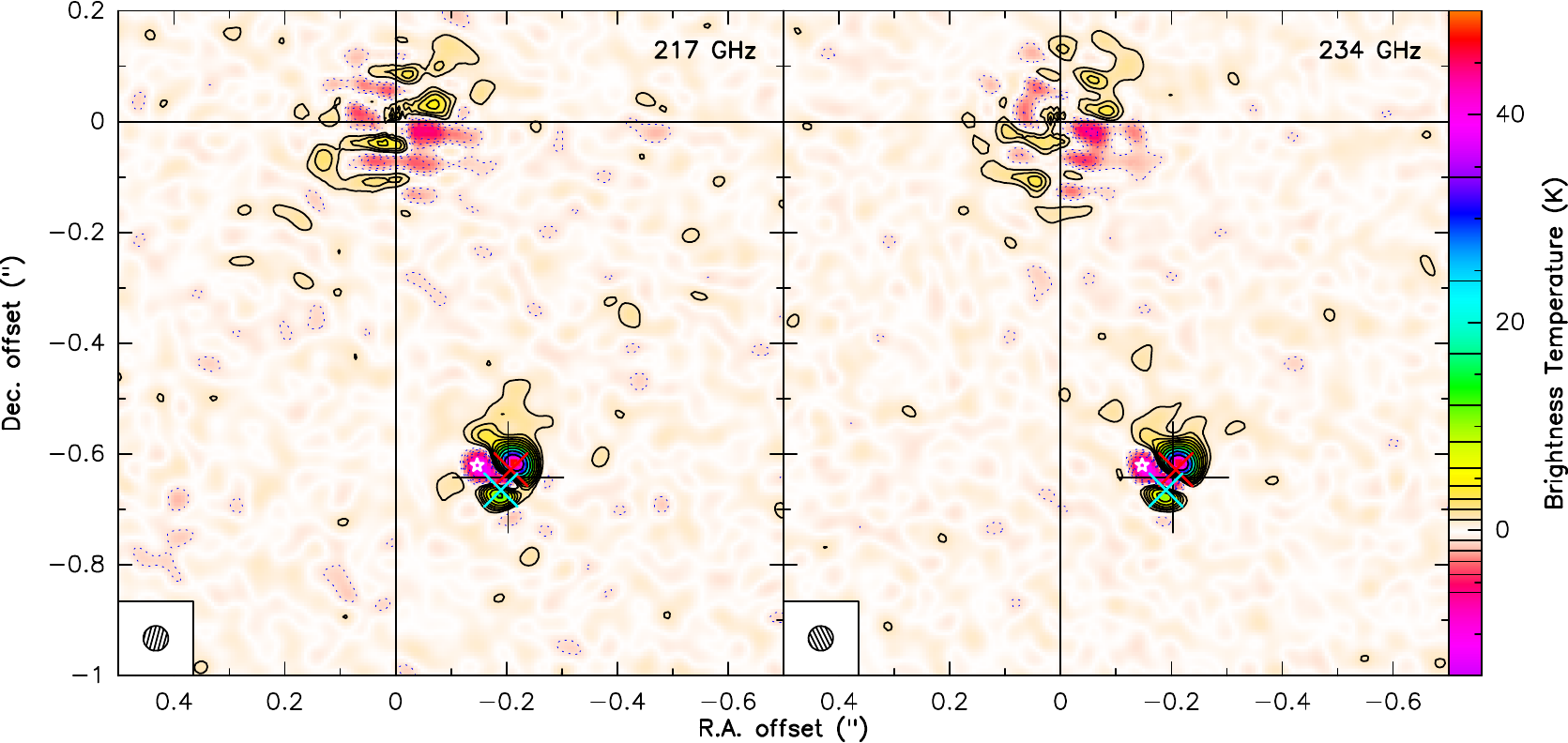}
\caption{Difference between the two epochs (09-Nov-2021 subtracted from 23-Jun-2023),  at 217 GHz (left) and 234 GHz (right). The positive regions in these images reveal where the emission is enhanced in 2023 with respect to 2021. Red crosses show the positions of T Tau Sa, and cyan show the positions of Sb; both are for the 2023 epoch.  A small white star shows the position of Sb in 2021 and is coincident with a negative subtraction residual.  The black plus is the position of the Sa/Sb center of mass. Beam sizes are presented at lower left for clarity.
\label{fig:fig4}}
\end{figure}

Fig.\ref{fig:fig4} shows the difference in emission in the 217 (left) and 234 GHz (right) images from 2021 and 2023.
X's are as in Fig.\ref{fig:fig3}, and the white star is the position of the emission near Sb in Aug 2021, which thus appears as negative signal in this 2023-2021 difference. 
The images are in brightness units (K, with 1\,K = 77.8 $\mu$Jy/beam at 217 GHz, and 90.8 $\mu$Jy/beam at 234 GHz)

From these difference images in Fig.\ref{fig:fig4} and the strongly enhanced emission seen to the northwest of the binary, it is clear that the periastron passage has significantly disturbed the disk around Sa. 
Moreover, the flux from the surroundings of Sa and Sb has increased by a factor about 45\%, far beyond the 1.5\% reproducibility of the T Tau N flux. 
Emission from Sb is known to be variable and in part of non-thermal nature, so a detailed accounting of the origin of the variation is somewhat hampered by the lack of clear spatial separation between Sa and Sb in the matched resolution maps. 
However, the measured peak flux towards Sa increased in 2023 by about 38$\pm3$\%. This mean and uncertainty range is derived from two different measurements made using slightly different areas (both avoid the flux from T Tau Sb).  So, the dominant fraction of the flux increase in the environment of T Tau S is from Sa.  This substantial increase in mm flux likely predominantly results from heating of the Sa disk in response to enhanced stellar accretion activity triggered by the close periastron passage of Sb \citep{beck20}.  Additionally, the flux change is about 2.5\% larger at 217 GHz than at 234 GHz. 
This small difference at the two frequencies could be linked to the spatial extent of the source and different angular resolutions of the measurements, yet T Tau N does not exhibit this behavior.


\subsection{350 GHz Measurements}
\begin{figure}
\includegraphics[width=18.0cm]{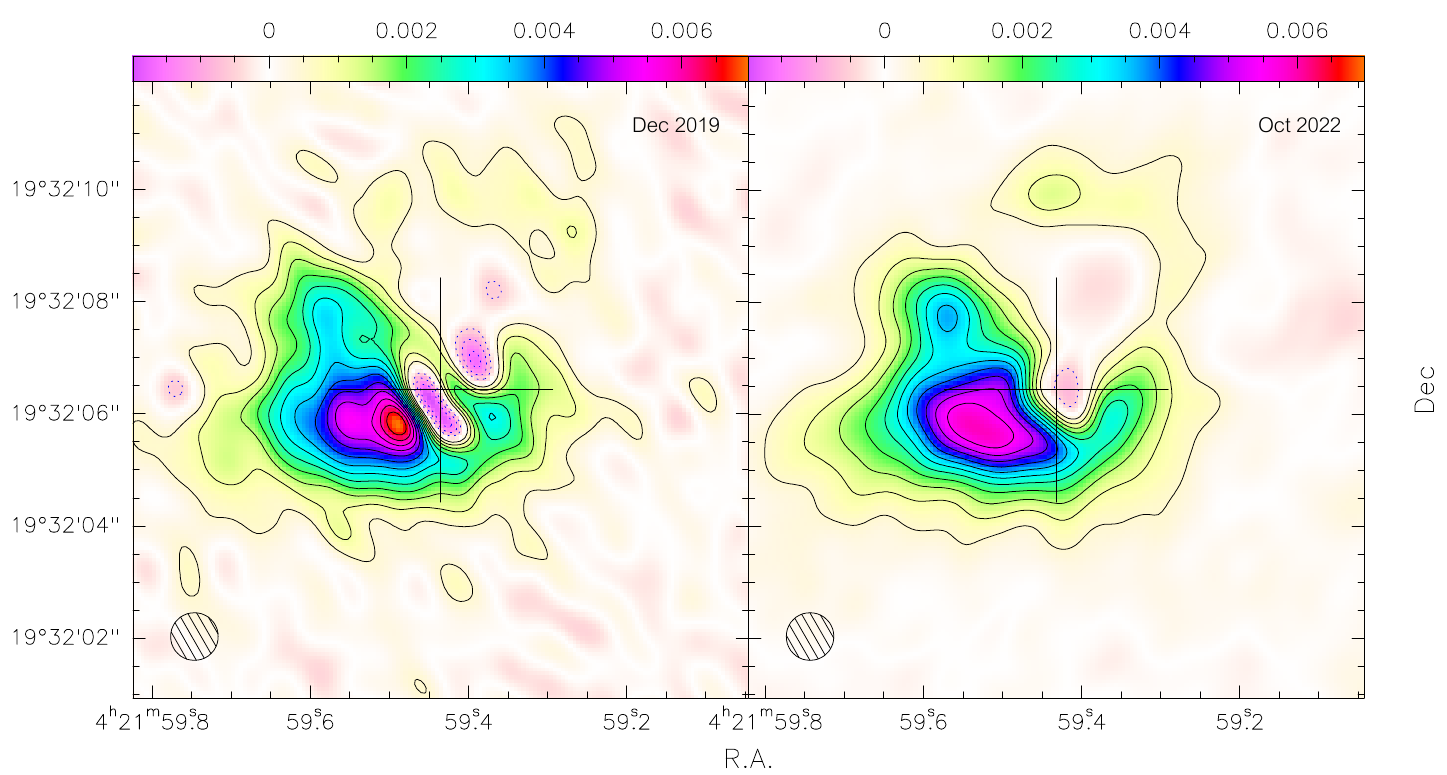} 
\caption{
From low resolution 356 GHz measurements ($\sim$0.7\farcs; Table 1), the residual emission around T Tau in Dec-2019 (left) and Oct-2022 (right)
after removal of N, Sa and Sb source models. Spatial resolution is 0.85$''$. Contour levels are 7 mK, or approximately 
0.5 mJy/beam, corresponding to 2.2 and 5.8 $\sigma$ resp in Dec-2019 (left) and Oct-2022 (right). The large black cross is the T Tau N position. Beam sizes are presented at lower left for clarity.
\label{fig:fig5}}
\end{figure}

The 350 GHz data was obtained at two different epochs; 2019 and 2022. The moderate 
angular resolution data from 2019 (about 0.8$''$ with proper weighting and tapering) precludes
a clear separation between T Tau N and Sa/Sb. 
Using the fixed positions and sizes determined from the high resolution observations at 225 GHz, and further assuming that the flux from Sb has a flat spectrum, we derive the flux densities
for the N and Sa disks quoted in  Table \ref{tab:350} from a fit in the $uv$ plane using
all available baselines (up to about 380 k$\lambda$).
Low resolution ($0.85''$) images of the fit residuals are presented in Fig.\ref{fig:fig5}, showing the extended emission.

The measured fluxes of T Tau N at 350 GHz are consistent with a
non variable source, and a spectral index of 2.1 when 
compared to the measurements at 225 GHz.   Time variability of T Tau S is directly visible at 350 GHz. Assuming T Tau N flux was constant, the flux of Sa has increased by 8\% between the two epochs. The faint extended structure surrounding the system also shows a similar increase (10\%, see Table \ref{tab:350}), but no conspicuous morphological changes (see Fig.\ref{fig:fig5}).
Additionally,
the spectral index of Sa between 225 GHz and 350 GHz varies
between 2.8 and 2.5, higher than that measured at 225 GHz (2.0).
This discrepancy may have two origins. First, it is possible that some extended flux may contaminate the
much lower angular resolution 350 GHz measurements, despite
the fixed sizes and locations used in the derivation. Secondly, the source is definitely time variable (Sa brightened by about 23 \% at 225 GHz between Nov 2021 and Jun 2023), and the spectral index may appear to change because the 350 GHz and 217/234 GHz measurements are not simultaneous. 

\begin{table}
\begin{tabular}{c|c|c|c|c}
\hline
 Date & Flux N & Flux Sa & Flux Sb &  Extended \\
       & (mJy) & (mJy) &  & \\
\hline
Dec 2019 
 & $    456.3  \pm     0.1$ 
 & $     24.7  \pm     0.1$ 
 & $     [1.4 ] $ 
 & $     60 \pm 1 $ \\
Oct 2022 
 & $   450.0 \pm      0.1$ 
 & $    26.4  \pm     0.1$ 
 & $     [2.1 ] $ 
 & $  65 \pm 1 $ \\
\hline
\end{tabular}
\caption{350 GHz flux densities of T Tau N, Sa and Sb}
\label{tab:350}
\end{table}

\subsection{Sa/Sb Positions and Orbital Motion}

To better characterize the Sa / Sb system, we removed in the $uv$ plane emission from the T Tau N circumstellar disk in the high resolution 225GHz observations, and fitted the residual by two elliptical Gaussians. 
Two methods were used to remove T Tau N emission:  subtraction of the Clean components, and removal of the source model made of a uniform (inclined) disk and an elliptical Gaussian. 
Both methods yield similar fit results. 
Only the longest baselines (above 500 m, or 450 k$\lambda$) were retained for the fit, to avoid being biased by the faint extended emission around Sa/Sb. 
Results are given in Tables \ref{tab:sources2023} and \ref{tab:sab}. 
The Sa disk flux increased by 25\% between the two epochs while that of Sb increased by $50 \pm 5$\,\%.   
The derived positions were used for Fig.\ref{fig:fig4} and \ref{fig:fig7}.  
The separation of T Tau Sa and Sb in the 2023 epoch was just 41.3 milli-arcsec, or 5.7~au, based on the distance to T Tau.  These ALMA measurements have been particularly important for the orbit modeling because as T Tau Sb approached periastron, it has been very faint in the near IR and difficult to measure in ground-based adaptive optics imaging.
In the two years separating our high resolution ALMA measurements, T Tau Sb moved by $\sim$80 degrees in position angle relative to T Tau Sa, and the separation between the two stars decreased by 8 milli-arcsec, or about 1.2~au,  The time frame sampled by these two measurements encompassed the $\sim$March 2023 periastron passage.  The location predicted by the orbit fit in \citet{scha20} gives a separation of 36.7 mas at a position angle of 149.1$^\circ$, which is a discrepancy of 5.9 mas (0.9 au) from our measured value.

\begin{figure}
\includegraphics[width=12.0cm]{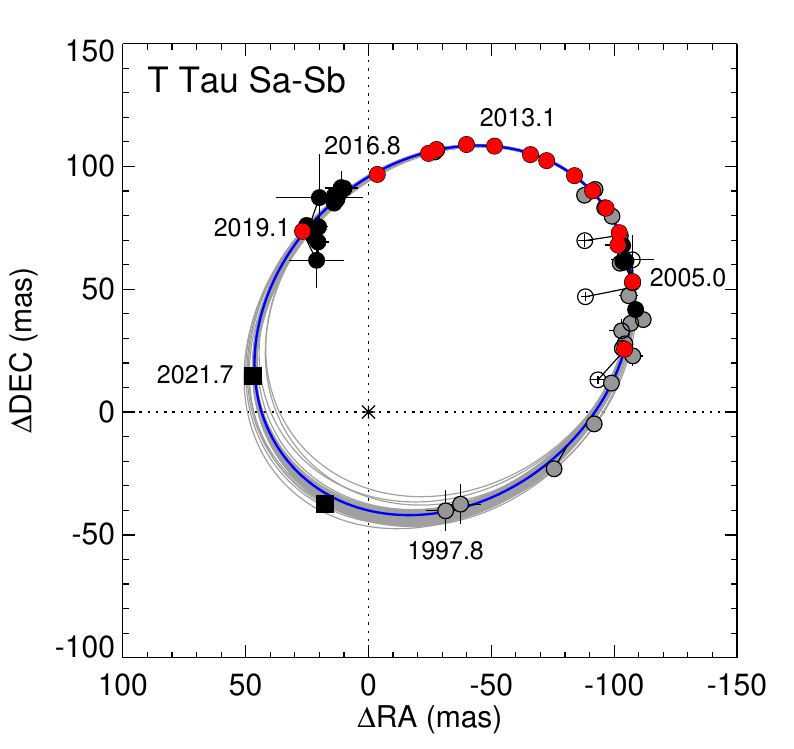}
\caption{
  The new orbit model (blue curve) that incorporates the new ALMA measurements of T Tau Sb position with respect to Sa in 2021 and 2023 (seen as black squares).  The gray curves present the range of model fits from the ensemble sample of orbits used to estimate uncertainty.  The other symbols are as in \cite{scha20}:
red circles are Keck NIRC2 adaptive optics (AO) measurements as presented in \cite{scha20}, the black circles are Gemini NIRI AO measurements from our team \citep{scha20,beck20} and the gray circles are published values from the literature \cite{scha20}. 
}
\label{fig:fig6}
\end{figure}

A new orbit fit to the infrared positions used in \cite{scha20} and the ALMA positions in Table \ref{tab:sab} was carried out.  This new fit is presented in Figure \ref{fig:fig6} and gives the parameters outlined in Table \ref{tab:orbitfit}.  We computed uncertainties through a Monte Carlo bootstrap approach, where we randomly selected measured positions from the sample with repetition. We then randomly varied the measurements within their uncertainties and refit the orbit. We repeated this process 10,000 times and adopted uncertainties from the standard deviation of the bootstrap distributions. 
Based on the best fit and this sample of orbits used to determine the uncertainties, the predicted separation of T Tau Sb in 2023 is 37.5 $\pm$ 1.8 mas at a position angle of 154.6$^\circ$ $\pm$ 8.9$^\circ$. The measured separation in 2023 is discrepant from the predicted position by 3.8 $\pm$ 1.9 mas (0.55 au). The fact that the predicted position from the previously computed orbit and the updated orbit is smaller than the measured value might indicate that there is extra extended emission that is biasing the ALMA position in 2023. 

\begin{table}
\begin{tabular}{l|l|l|l|l}
\hline
 Date & Flux Sa & Flux Sb & Separation &  PA  \\
       & (mJy) & (mJy) & (mas) & $^\circ$ \\
\hline
Nov 2021 
 & $    7.25 \pm     0.02$ 
 & $    1.37 \pm     0.02$ 
 & $     49.2 \pm      0.3$ 
 & $     72.7 \pm      0.3$ \\ 
Jun 2023 
 & $    8.96 \pm     0.04$ 
 & $    2.07 \pm     0.04$ 
 & $     41.3 \pm     0.5$ 
 & $    154.6 \pm     0.1$ \\ 
\hline
\end{tabular}
\caption{Sa / Sb flux densities at 225 GHz and angular separation}
\label{tab:sab}
\end{table}

\begin{table}
\begin{tabular}{l|l|l}
\hline
Orbit Model & Best Fit & Uncertainty \\
Parameter & Value &  \\
\hline
$P$     &     27.29   &     0.40  \\
$T$     &     1995.87    &    0.35  \\
$e$      &   0.547   &   0.017  \\
$a$      &    84.36    &    0.77  \\
$i$     &     17.31     &    3.39  \\
$\Omega$    &      95.20    &     16.50  \\
$\omega$     &     43.60    &     17.54  \\
M$_{\rm tot}$ \Msun{} & 2.65  & 0.12 \\
\hline
\end{tabular}
\caption{Updated Orbital Parameters for T Tau Sa,Sb}
\label{tab:orbitfit}
\end{table}

\subsection{A Closer Look at T Tau N}
\label{sec:ttaun}

\begin{figure}
\includegraphics[width=18.0cm]{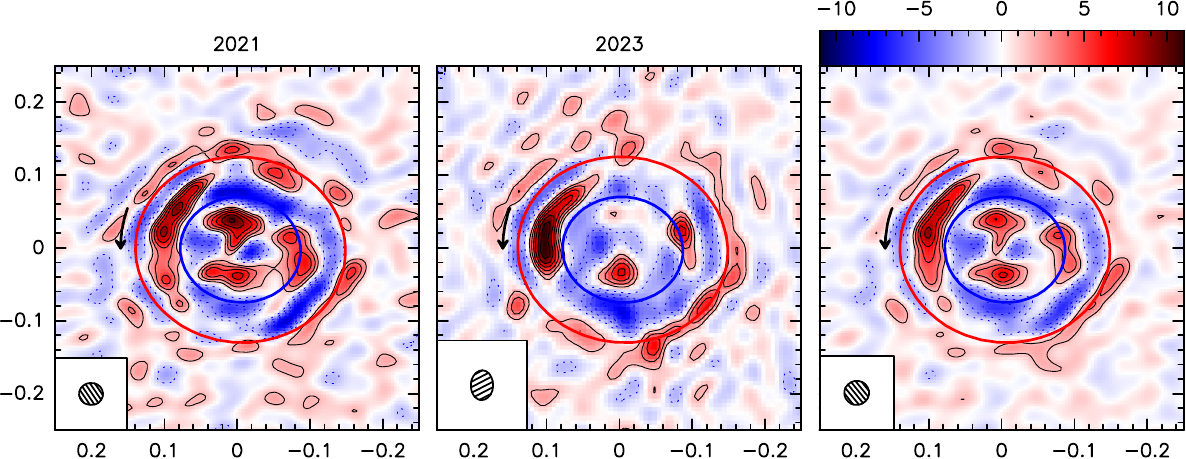}
\caption{
  Emission from the disk of T Tau N after subtraction of the Gaussian and uniform inclined disk model. The red ellipse indicates the uniform disk edge, and the blue one the position of the gap from \citet{yama21}. The black arrow indicates the expected Keplerian rotation of the bright elongated structure.
  Left panel is for Jun 2021, middle panel Oct 2023, and right panel is the 
  noise weighted average. Contours are in steps of 1.7\,K, about $3\,\sigma$ for each epoch, and
  $3.5\,\sigma$ in the combined image. The color scale is for brightness in K.
  \label{fig:fig7}
}
\end{figure}

We showed in Section \ref{sec:sub:images} that the mm flux model for T Tau N was well fit by a Gaussian central source superimposed on a uniform disk.  However, subtracting the model from Table\,\ref{tab:sources} for T Tau N from the images reveals residual structure that is more complex than the simple uniform disk.  Fig.\ref{fig:fig7} presents the model subtracted residuals for the 225 GHz emission in the 2021 epoch, 2023 epoch and a noise-weighted average image.  The overall morphology of the residual emission seen in the model-subtracted images of T Tau N is similar at both epochs, though azimuthally asymmetric structures may vary slightly because of disk Keplerian rotation.  Evidence that the disk edge is smoother than assumed by the simple uniform model is seen as weak emission detected outside the uniform disk radius and a small deficit of emission just inside the edge.  Within the uniform disk region, the residuals can roughly be divided into 3 features (Fig.\ref{fig:fig7}):

{\bf(a)} a deficit of emission about 80~mas (12~au) from the center, 

{\bf(b)} a strong, crescent-shaped and elongated region of excess emission at PA 20-90°, and 

{\bf(c)} point source enhancements in emission 40~mas North and South of T Tau N, about 6.6\,au deprojected distance from the star.

The observations presented here naturally have a linear resolution of about 4.5 au, similar to that obtained by \citet{yama21} through the use of a super-resolution sparse modeling technique \citep{naka19}. 
The relative emission minimum (feature a) at $\sim$12~au matches in position the gap also found by \citet{yama21}. 
The overall contrast between the 12~au gap region in the disk of T~Tau~N and the surrounding material is at a level of about 12-15\,K (while the uniform disk model has a brightness of 70\,K).  This is also consistent with the findings of \citet{yama21}.
\citet{yama21} suggested that the gap could be created by a 0.5 to 3 Saturn mass planet, but we find no obvious point source within this gap that would correspond to a circumplanetary disk around such a planet.

Our observations reveal a complex morphology in the disk of T Tau N.  For example structure (b) described in the previous paragraph looks akin to the crescent-shaped features seen in the disks of other sources such as AB\,Aur \citep{tang17}, and MWC\,758, HD\,142527 \citep{bae23}.  The centroid of the crescent-shaped feature in the disk of T~Tau~N is consistently located at a distance of $106 \pm 3$\,mas, 15.4 au from the disk center\footnote{This distance is 100 mas if one uses instead the GAIA position of T Tau as center.}.   At this distance, using a mass of 2.03 \Msun{} for T Tau N \citep{scha20}, the Keplerian orbital period is 42 years. There is an apparent displacement of crescent-shaped structure between our two observation epochs (seen as the arrow in \ref{fig:fig7}) which is in the expected direction if the disk around T Tau N rotates in the same direction as the Sa-Sb orbit. Morphological changes above the 3 $\sigma$ level between the two epochs make it challenging to measure a precise proper motion. Fig.\ref{fig:fig7} suggests an angular displacement by $\sim 20^\circ$, in approximate agreement with the orbital motions over the 26 months lapse time, given a 42 yr Keplerian period at this distance from the star.  Proper motions of similar features in MWC\,758 were reported by \citet{kuo24}, who concluded that a non-Keplerian orbital speed was ruled out in a vortex model often invoked to explain such azimuthal asymmetries \citep[e.g.][]{baru19}.  The T Tau N disk differs from other sources with azimuthal structures in that both its spectral index and apparent brightness clearly indicate the dust is optically thick at 225 GHz.  As a result, the measured brightness variations may be less sensitive to surface density contrasts than to dust temperature differences that may result from illumination changes. It is worth noting that in the case of vertical temperature gradients, and provided dust settling is not too strong, surface density variations will result in apparent brightness contrast even in the optically thick case. 

The origin of the compact structure(s) (c) described above remains obscure. The orbital period at this radius is about 12 yr, Keplerian rotation of real dust structures between the two epochs would amount to a $65^\circ$ shift in position angle between the two epochs. The observed changes in feature (c) have no relation to this. They may reflect illumination changes occurring on much shorter timescales, and sparsely sampled at the two observing dates.

In summary, the disk of T Tau N shows internal structure in the optically thick dust.  Gap carving by hidden planets forming within the disk could be a cause for the observed sub-structure.  However, it is also feasible that the observed structures in the disk of T~Tau~N could be induced by the gravitational influence of the binary companion Sa-Sb, which is modulated by the 27 yr period of the Sa-Sb orbit. Theoretical simulations show that warps and substructure within an optically thick disk can be caused by gravitational interactions through stellar flybys in bound or unbound orbits \citep{ostr94, larw96, cuel18, kurt24}.  As the perturbing star goes through periastron passage, spiral arms can be excited in the primary star disk \citep{smal23}.  Measurement of significant structure now in the disk of T Tau N would be consistent with a model where the wide orbit of T~Tau~N and S have recently undergone periastron passage, as is thought to have occurred based on current orbital models \citep{scha20,kohl16}.  Interaction between T~Tau~N and the S binary occurs on long ($> 1000$ yr) timescale due to the wide N-S orbit, and flyby induced structure is theorized to decline following closest approach \citep{smal23}.

\section{Discussion}

\begin{figure}
\includegraphics[width=18.0cm]{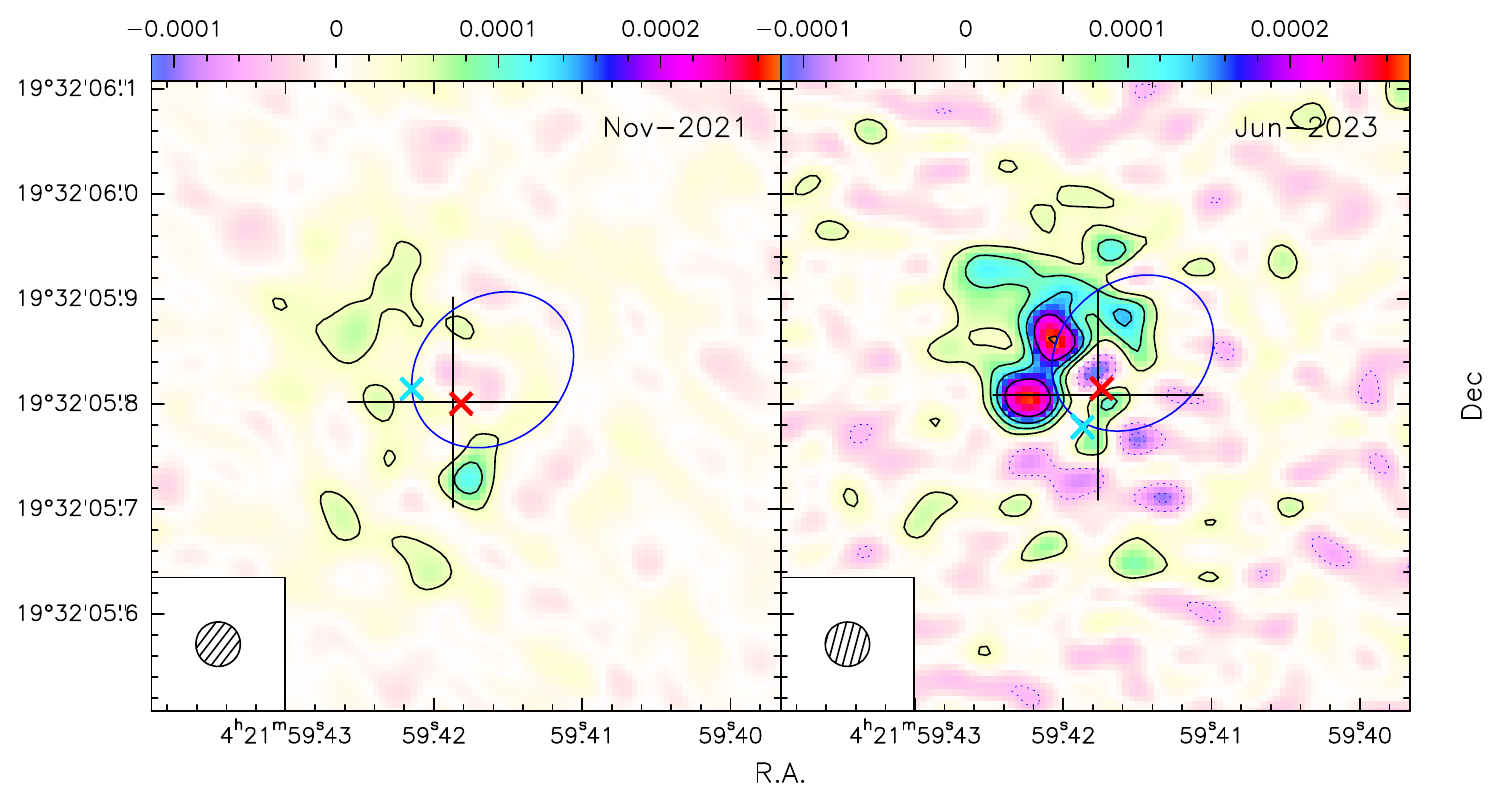}
\caption{
  Residual 225 GHz emission after subtraction of the best fit model 
  of the Sa and Sb compact sources.
  Contours are in steps of 45 microJy/beam, or 0.6 K, corresponding
  to 4.4 $\sigma$ and 2.8 $\sigma$ respectively for 2021 and 2023.  Red and cyan X's trace the T Tau Sa, Sb positions, respectively.
  The blue ellipse is the Sb orbit relative to Sa as in Fig.\ref{fig:fig1}. Beam sizes are presented at lower left for clarity. The red and cyan X’s mark the Sa and Sb positions, respectively.
}
\label{fig:fig8}
\end{figure}

The T~Tau system is an important laboratory to investigate dynamical effects of stellar periastron flybys because the full orbit of Sa~+Sb can be monitored on a reasonable time scale and the stars and their disks can be studied in detail during periastron passage.  Simulations show that stellar flyby encounters can disrupt and heat the disks, strip material off of the primary circumstellar disk, and the perturber itself could capture and accrete material through the interaction \citep{clar93}.  Figure \ref{fig:fig8} shows the residual 225~GHz emission after removal of the modeled Sa and Sb emission in the environment of T Tau S in the two observing epochs, 2021 before binary periastron passage and 2023 just afterward. Red and cyan `$\times$' show the modeled position of T Tau Sa and Sb, respectively.  In 2021, the residual emission is low level, and is greatest to the south west of T Tau Sa. In 2023, the residual emission brightened considerably, and the peak emission had shifted to the east of the close binary. The total flux around Sa and Sb in the 2023 residual is just $1.4\pm0.1$ mJy/beam. This is unfortunately too small to derive an accurate spectral index between the 217 and 234~GHz observed frequencies. The total flux in a similar region in Nov 2021 is about 0.5 mJy. The increase in emission that appeared between the two epochs is about 10\% of the Sa disk flux and is detected at a high level of signal-to-noise.

Without a clear knowledge of the spectral index of this increased emission in the environment of T Tau S in June 2023 following the periastron passage, we unfortunately cannot firmly conclude its origin.  However, there are two likely scenarios.  
First, it may be due to thermal emission from dust gravitationally disrupted and removed from either the Sa or Sb disk, or both.  
In this scenario, the excess flux can be directly converted into a corresponding dust mass, assuming optically thin emission. Using an absorption coefficient of 0.02 cm$^{2}$\,g$^{-1}$ and a dust temperature of 30 K, the minimum (dust+gas) mass in this material is $\sim10^{-4}$ \Msun\, \citep{dutr96}. By comparison, the Sa disk is at least 10 times more massive, and most likely much more since its emission appears to be optically thick.

The second likely possibility is that the excess emission could be of non-thermal origin, driven by Sb. 
Such non-thermal mm emission could arise from energetic electrons interacting with the disturbed magnetic field of T Tau Sb as the star passed very close to Sa through the periastron passage. 
The location of the excess emission along the apparent Sb orbit may support the theory that the increased emission in the 2023 epoch observation arises from such non-themal processes.  
T Tauri is known to have complex radio emission that is both variable and circularly polarized \citep{skin94, ray97}.  
The radio emission arises from non-thermal gyrosynchrotron radiation and has both compact and extended (to $\sim$1$"$) emission components \citep{skin94}.  
High-resolution measurements show that the compact non-thermal radio emission is stronger toward T Tau Sb compared to Sa \citep{john04}.  Nonthermal mm emission is not commonly seen in T Tauri stars, except in the most energetic and magnetically active systems such as the binary eruptive variable DQ Tau \citep{salt10}. Time variable non-thermal emission extending up to the mm regime has also been detected in the the V\,773\,Tau multiple system \citep{dutr96}, and attributed to synchrotron emission driven by periastron passage in the close binary V\,773\,Tau\,A \citep{mass02,mass06}.  The slight mis-match between the best-fit position of the mm emission from T Tau Sb and the orbit model fit, as discussed in the previous section, supports the interpretation that this emission is non-thermal.  The 3.8 mas discrepancy in position is 0.53 au, which could be caused by non-thermal emission from an outflow, or magnetic field structure that is spatially extended from the position of the Sb stellar surface and in the opposite direction from Sa.  An accurate measurement of the mm spectral index of extended emission in the immediate environment of the T Tau S binary at high resolution across the 225 and 350~GHz bandpasses could show if this is clearly non-thermal emission, or if we have witnessed tidal disruption of dust in the disks triggered by the close binary periastron passage.  

While the nature of the change in mm emission in the environment around T Tau Sb is uncertain, it is clear that the increased flux detected in the large (arcsecond) scale structure
East of Sa/Sb measured at 350 GHz (Fig. \ref{fig:fig5}) is due to an overall warming of the dust. The measured increase in this emission is consistent with the similar change found for the optically thick dust disk around Sa (Fig.\ref{fig:fig4}). Hence, there is clear evidence for heating of dust in the system between the two epochs of our observations.  The dust heating is likely from a combination of predominantly environmental heating through stellar accretion activity, which is known to increase near periastron passage for T Tau S \citep{beck20}, and dynamical energetic effects.  The origin of the overall mm flux changes in the environment of T~Tau~S during the recent orbital periastron passage is likely a combination of both non-thermal magnetic and thermal dust heating processes.
 
\section{Summary}
In this study, we present new measurements of the T Tauri triple system observed with ALMA across two epochs.  In Summary:

1) The T Tau Sa / Sb binary is clearly resolved in our two epochs of ALMA measurements spanning the recent March 2023 periastron passage of Sb around Sa.

2)  In the two epochs of observations around periastron passage, we find clear increases of millimeter  emission in disk around T Tau Sa and the overall environment around the binary.  The spectral indices show that this is largely thermal dust, the heating is likely triggered by the orbital motion and increased accretion activity in T Tau Sa during periastron passage.

3)  The morphology of extended mm emission in high resolution maps has changed in the two epochs of observations; it has increased in magnitude and extent in the environment around T Tau Sa-Sb.  This enhanced extended emission could be caused by non-thermal processes or by tidal disruption of dust in the disks.  Measurement of more accurate spectral indices could distinguish between these two scenarios.

4)  Our high resolution ALMA observations provide two new measurements of the position of T Tau Sb relative to Sa, which allows us to update orbit models for the system. 

5)  We measure structure in the optically thick disk of T Tau N, including a 12~au ring gap in the disk material and a crescent shaped region of enhanced emission at 15.4~au which appears to be moving in Keplerian orbit between our two measurement epochs.

\begin{acknowledgments}
TLB and all of the authors posthumously thank Michal Simon, advisor, mentor and friend, for his tireless support and enthusiasm for our research efforts.  Rest in peace, Mike.  You will be missed.  This paper makes use of the following ALMA data: ADS/JAO.ALMA\#2019.1.00703.S and ADS/JAO.ALMA\#2022.1.01134.S. ALMA is a partnership of ESO (representing its member states), NSF (USA) and NINS (Japan), together with NRC (Canada), NSTC and ASIAA (Taiwan), and KASI (Republic of Korea), in cooperation with the Republic of Chile. The Joint ALMA Observatory is operated by ESO, AUI/NRAO and NAOJ.
\end{acknowledgments}

%

\vspace{5mm}
\facilities{ALMA}


\software{astropy \citep{2018AJ....156..123A},  
          casa \citep{casa22},
          Imager \citep{25a} Interferometric Imaging package developed at LAB/OASU}







\bibliography{beck25}{}
\bibliographystyle{aasjournal}



\end{document}